\documentclass[showpacs,preprintnumbers,amsmath,reprint,aps,prl]{revtex4-1}

\usepackage{bm}
\usepackage{graphicx}
\usepackage{color}
\usepackage{amsmath}
\usepackage{epstopdf}
\usepackage{upgreek}

\newcommand{\commentOut}[1]{}

\begin{document}

\title{Magnetic quantum number resolved state-to-state chemistry }

\author{Joschka Wolf}
\author{Markus Dei{\ss}}
\author{Johannes Hecker Denschlag}

\affiliation{Institut f\"{u}r Quantenmaterie and Center for Integrated Quantum Science and
Technology IQ$^{ST}$, Universit\"{a}t Ulm, 89069 Ulm, Germany}

\date{\today}

\begin{abstract}
We extend state-to-state chemistry to a realm where
besides vibrational, rotational and hyperfine quantum states
magnetic quantum numbers are also resolved. For this, we make use of the Zeeman effect which energetically splits levels of different magnetic quantum numbers. 
The chemical reaction which we choose to study is three-body recombination in an ultracold quantum gas of $^{87}$Rb atoms forming weakly-bound Rb$_2$ molecules. Here, we find the propensity rule that the total $m_F$ quantum number of the two atoms forming the molecule
is conserved. 
Our method can be employed for many other reactions and inelastic collisions and will allow for novel insights into few-body processes.
\end{abstract}

\maketitle
A primary goal of state-to-state chemistry is to gain an understanding about chemical reactions on a fundamental level (see, e.g. \cite{Nesbitt2012}).
This includes investigating what the final product states are
for a given quantum state of reactants.
 A proven experimental method for carrying out state-to-state chemistry is based on
 atomic and molecular beams (for recent reviews, see e.g. \cite{Yang2007, Jankunas2015, Meerakker2012}).
 In recent years, a complementary approach has been established,  investigating chemical reactions in traps in the ultracold regime (for reviews, see
\cite{Bohn2017, Quemener2012, Balakrishnan2016, Krems2008, Carr2009, Doyle2004}).

Within the last decades it has become standard for state-to-state chemistry experiments to resolve the electronic, vibrational and rotational quantum states of the molecular products. Very recently our group demonstrated that even hyperfine substates of the products can be resolved \cite{Haerter2013, Wolf2017}.  Magnetic substates, however, have  in general not yet been identified, apart from the special case  where the reaction is essentially only taking place in a single quantum channel. Typical examples for such single-channel dominated reactions occur in the vicinity of Feshbach resonances, e.g. \cite{Chin2010, Ferlaino2011, Knoop2010, Rui2017, Hoffmann2018}, or are photo-induced reactions, e.g. \cite{Jones2006, McDonald2018}.

\begin{figure}[t]
	\includegraphics[width=\columnwidth]{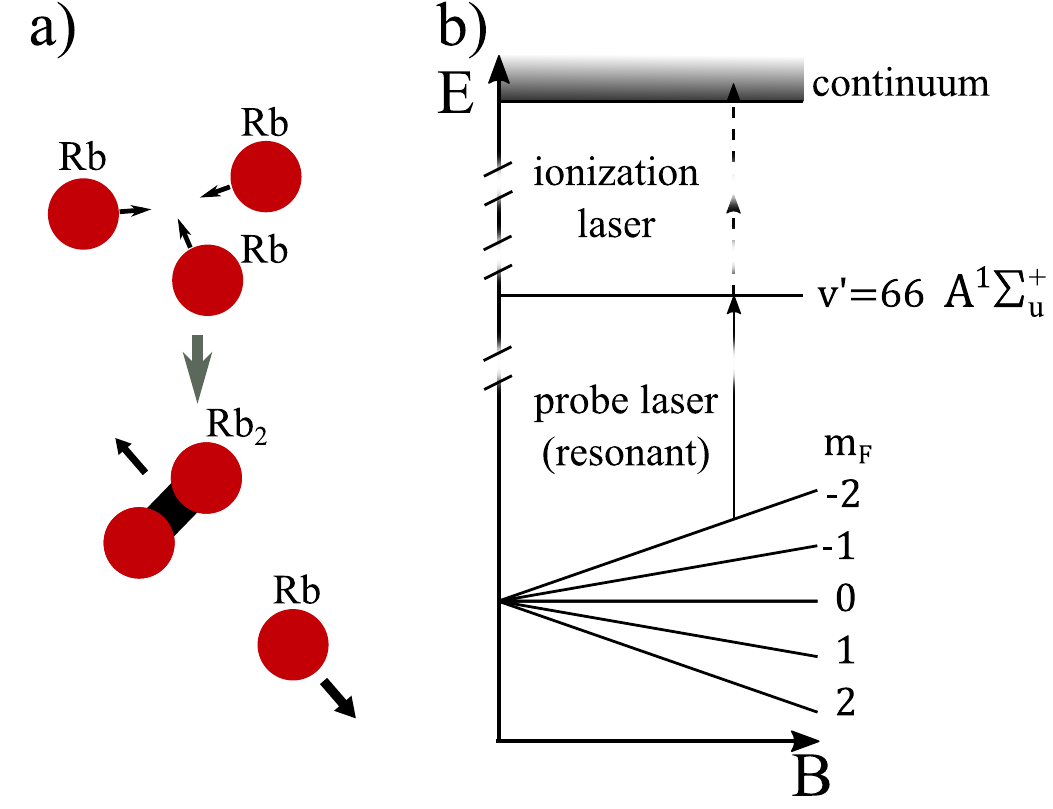}
	\caption{(a) Three-body recombination of three neutral Rb atoms forming  a Rb$_2$ dimer. (b) Schematic of the REMPI detection scheme, resolving the $m_F$ quantum number of the molecule. As an example we consider here a molecular level manifold with $F = 2$. The linear Zeeman effect due to a $B$-field splits the $F = 2$ levels into
		magnetic components $m_F=-2,-1,0,1,2$. The splitting is large enough so that each level can be resolved via resonant excitation towards the intermediate level $\text{v}'=66$, $A^1\Sigma_u^+$ with  the probe laser.
		From there, two photons of the ionization laser ionize the molecule. }
	\label{fig1}
\end{figure}

Clearly, state-to-state aspects become especially of interest when more than one reaction channel is  involved. Here, we extend state-to-state chemistry to finally resolve magnetic substates for systems where
many  reaction channels are present. This is achieved by measuring how the molecular product states are Zeeman shifted as a function of an applied external magnetic field. The chemical reaction which we choose to study is three-body recombination of $^{87}$Rb
where two atoms combine to form a Rb$_2$ dimer and the third Rb atom (spectator) carries away part of the released binding energy, see Fig.~\ref{fig1}(a). We work at low magnetic fields of a few G where atomic interactions are non-resonant \cite{Haerter2013, Wolf2017}.
We probe molecular quantum states in a range of small binding energies and find the following "no-spin flip" propensity rule. The $m_F$ quantum number of the molecule equals the sum of the initially prepared $m_f$ quantum numbers of the two atoms which form the molecule. Here, $m_F$ is the magnetic quantum number of the total angular momentum $F$ of the molecule excluding mechanical rotation and $m_f$ is the magnetic quantum number of the total angular momentum $f$ of the atom. The "no-spin flip" propensity rule indicates that the spectator atom participates in the reaction merely via mechanical forces, so that it induces no  spin-flips on the forming molecule.

We carry out our experiments with an ultracold, thermal cloud of $^{87}$Rb with a temperature of about $750\:\text{nK}$. The cloud consists of  $N_\text{at}\approx 5\times10^6$ atoms which are trapped in a far-detuned, crossed optical dipole trap
with a wavelength of about 1064$\:$nm. The trapping frequencies are $\omega_{x,y,z}= 2\pi\times (23,180,178)\:\textrm{Hz}$, where $z$ represents the vertical direction.
The atoms are in the electronic ground state and initially spin-polarized in the hyperfine state $f=1, m_f = -1$. From this state also other $m_f$ spin polarizations can be prepared.

Within the atomic gas, three-body recombination produces weakly-bound Rb$_2$ molecules in the coupled $X^1\Sigma_g^+-a^3\Sigma_u^+$ potentials.
Given the initial peak particle density $n_0\approx 0.9 \times 10^{14}\:\textrm{cm}^{-3}$ and a three-body recombination loss rate constant of
$L_3=(4.3\pm1.8)\times 10^{-29}\:\text{cm}^6\text{s}^{-1}$ \cite{Burt1997}, about $1.5\times10^5$ atoms are lost due to three-body recombination in the first $500\:$ms. 
The  molecular products are distributed over a range of different states, characterized by sets of vibrational, rotational, various spin and magnetic quantum numbers.
We state-selectively probe the molecular products via resonance-enhanced multi-photon ionization (REMPI), followed by ion detection, similarly as in \cite{Wolf2017}.
Figure $\:$\ref{fig1}(b) shows our REMPI scheme.  The probe laser resonantly drives a transition to an intermediate level,
which is located in the electronic state $A^1\Sigma_u^+$,  has  vibrational quantum number  $\text{v}'=66$ and a total angular momentum (excluding nuclear spin) $J'$. Hyperfine splittings of this excited level are negligibly small ($<3\:\text{MHz}$), such that the  $J'$-states form a simple rotational ladder with rotational constant $b_v = 443\:$MHz  \cite{Deiss2015, Drozdova2013}.
From the intermediate level, another two photons of the ionization laser  ionize the molecule. We use the dipole trap laser as ionization laser. Its optical frequency is about $281.62915(6)\:\textrm{THz}$. The probe laser and the dipole trap laser are wavemeter locked with a precision of a few MHz. The probe laser intensity is $\approx3\:\textrm{W\,cm}^{-2}$  at the location of the atoms. This is sufficient to excite the probed molecules  with high probability to the intermediate $A^1\Sigma_u^+$ level.

The ion detection works as follows. Directly after its ionization the molecule is trapped in a Paul trap \cite{Setup} which is centered on the
atomic cloud. Subsequently, the micromotion-driven ion elastically collides with ultracold atoms \cite{Wolf2017, Haerter2013b}. This leads to loss of atoms from the shallow dipole trap which is measured by absorption imaging. The losses increase with the number of trapped ions.

In order to spectrally resolve  different $m_F$ quantum numbers we
apply a magnetic $B$-field of several G to induce linear Zeeman splitting. As an example, Fig.\:\ref{fig1}(b) shows the splitting of  a $F = 2$ molecular bound state.
In principle, $\vec{F}$ and the molecular rotation $\vec{R}$ couple to form the total angular momentum $\vec{F}_\text{tot}$. However, for the bound states considered here, this coupling is quite weak, corresponding to splittings between different $F_\text{tot}$ levels smaller than $1\:\text{MHz}$ \cite{Wolf2017}.
Therefore, already the $B$-field of several G efficiently decouples $\vec{F}$ and $\vec{R}$.
A level for given $m_F$ and $R$ quantum numbers is then essentially (2$R + 1$)-fold energetically degenerate, according to the manifold of the $m_R$ substates.

\begin{figure}[t]
	\centering
	\includegraphics[width=\columnwidth]{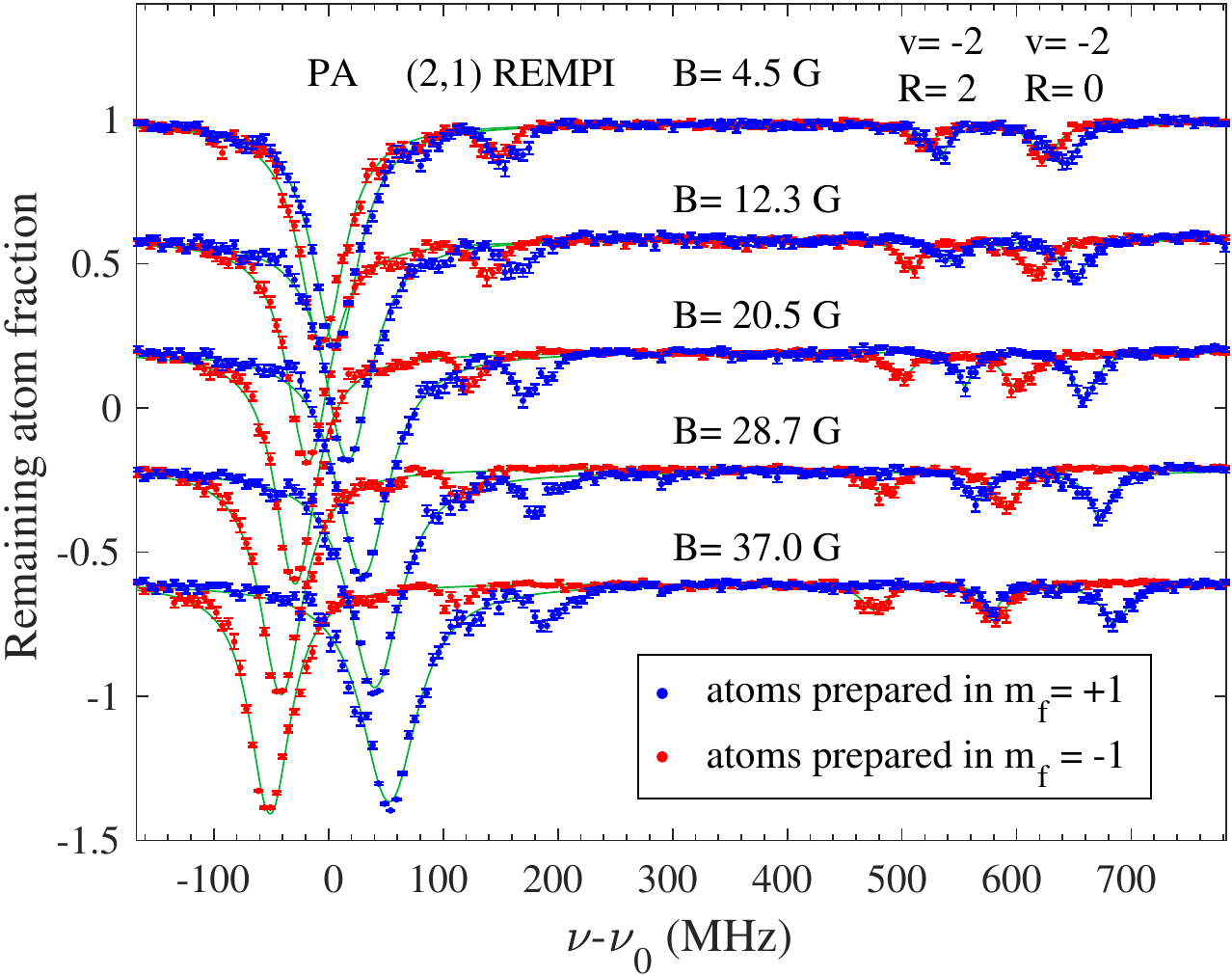}
	\caption{REMPI spectra for experiments where the atomic sample is initially prepared in the magnetic spin state $m_f=-1$ (red data) or $m_f=+1$ (blue data) for different magnetic fields $B$. Shown is the remaining atom fraction as a function of the probe laser frequency $\nu$.  $\nu_0=281,445.07(6)\:\text{GHz}$. Nominally the remaining atom fraction can range between 1 and 0, but
	in order to improve visibility data sets for different magnetic fields are offset relative to each other in vertical direction. The loss signals for $\text{v}=-2,R=0$ and $\text{v}=-2,R=2$ product molecules, the photoassociation line (PA), and (2,1) REMPI processes are labeled. The green lines are fits according to Eq.$\:$(\ref{eq:Lorentzian}) to the resonances. }
	\label{fig2}
\end{figure}

Figure \ref{fig2} shows several measured REMPI spectra as a function of the probe laser frequency $\nu$.
The  atoms of the cloud have initially been prepared either in state $m_f=-1$ (red data) or in state $m_f=+1$ (blue data).
 The measurements are carried out for various homogeneous  magnetic fields in a range between $4.5$ and $37.0\:\text{G}$.
 REMPI is carried out by switching on the probe laser for $500\:\text{ms}$.
Afterwards, the magnetic field is turned off. We measure the 
 remaining atom number  $160\:\text{ms}$ after the probe laser has been switched off, 
 so that ions  produced near the end of the REMPI period can still lead to atom losses. 
After normalization with the remaining atom number from an experiment where we work with  a far off-resonant probe laser we obtain the remaining atom fraction, as shown in Fig.\:\ref{fig2}.

We first discuss the data set for $m_f=-1$ and the lowest magnetic field of $4.5\:\text{G}$. The strong resonance line at $\nu- \nu_0 \approx 0$ is a photoassociation (PA) signal, where two $f=1,m_f=-1$ atoms which collide in an $s$-wave form a molecule in state $\text{v}'=66$, $A^1\Sigma_u^+$ with $J'=1$.
The observed width of the PA line is larger than the natural linewidth of about 12$\:$MHz (FWHM) of the excited level due to saturation effects caused by atom loss.

We now focus on the loss signals about $520\:\text{MHz}$ and $640\:\text{MHz}$ above the photoassociation line, which correspond to $\text{v}=-2,R=2$ and $\text{v}=-2,R=0$ product molecules, respectively. Here, $\text{v}$ is the vibrational quantum number and $R$ is the rotational quantum number. $\text{v}$ is counted downwards starting from $\text{v}=-1$ for the most weakly bound vibrational state correlated to the atomic $f = 1, f = 1$ asymptote \cite{LevelStructureInfo}. The $\text{v}=-2,R=0, 2$ molecular products have been detected and identified in previous work of ours \cite{Wolf2017}. Their $F$ quantum number is $F = 2$.
 Figure \ref{fig2} shows that their resonance lines shift towards lower probe laser frequencies when increasing the magnetic field $B$, in parallel to the photoassociation line.
Furthermore, we note that the depths and the widths of the lines stay the same when the $B$-field is changed. This suggests that
 the  molecules are produced in a particular  $m_F$ state, not a mix.
We observe a very similar behavior when we switch to experiments with $f = 1, m_f=+1$ atoms \cite{SpinFlipMethod} (blue data).
Now, however, all lines shift  in opposite direction as compared to the $f = 1, m_f=-1$ data.
The observed linewidths are  30$\pm10$ MHz (FWHM) which is a
factor of 2 to 3 larger than the natural linewidth. The increased width can be explained as the result of several factors, such as the laser linewidth, a Doppler shift of the moving Rb$_2$ molecules, and saturation effects.
We note  that besides the already discussed lines in Fig.\:\ref{fig2}, there are two additional lines (at around $\nu-\nu_0=90\:$MHz and 140$\:$MHz for the data sets at $B=4.5\:\text{G}$). These belong to so far unidentified molecular bound states which are ionized via (2,1) REMPI through a different intermediate state, see also \cite{Wolf2017}.

\begin{figure}[h!]
	\includegraphics[width=\columnwidth]{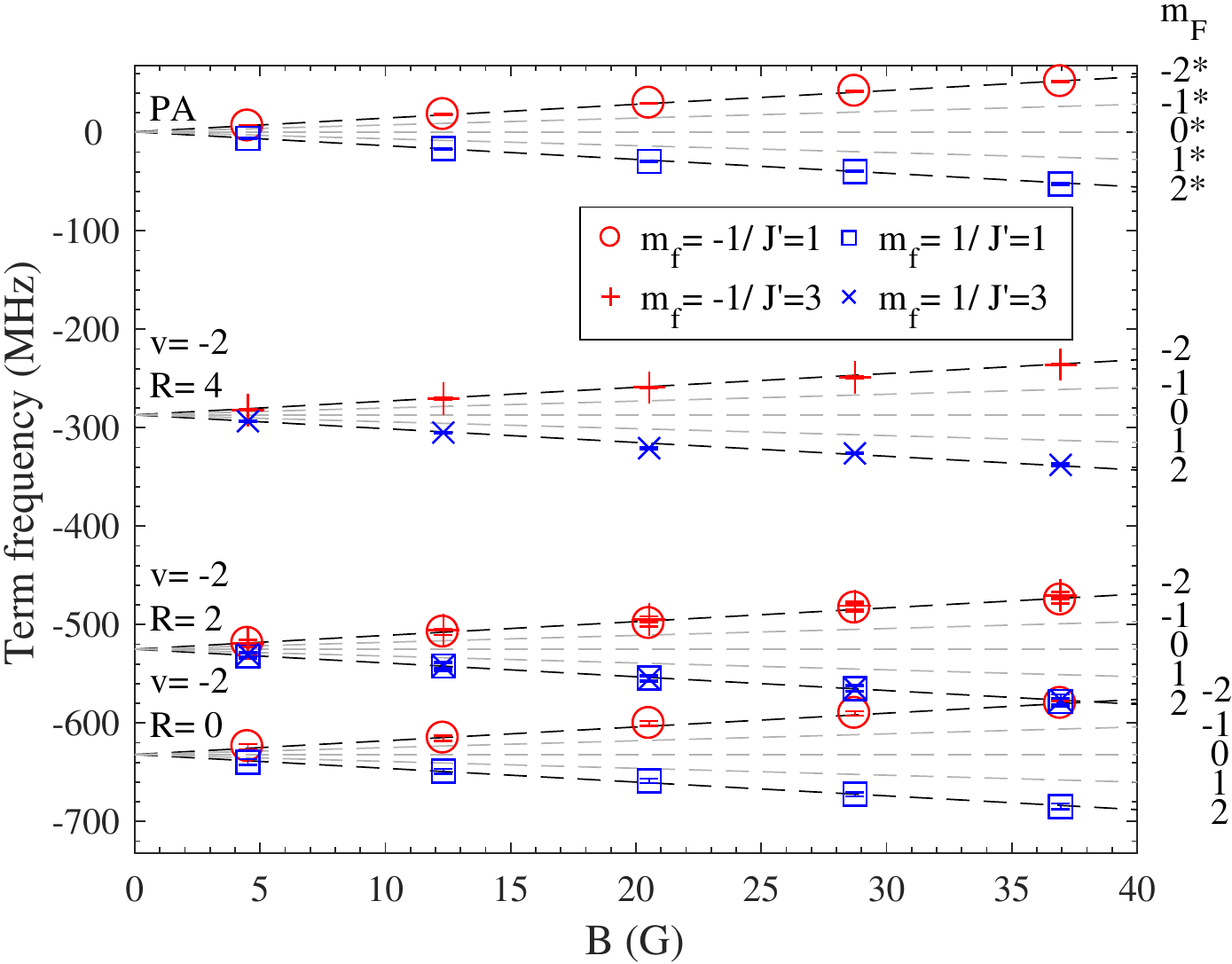}
	\caption{Zeeman shifts.  Shown are the term frequencies  (i.e. negative laser frequency positions relative to the threshold at $B = 0$) of the two-atom scattering states (PA) and the molecular levels $\text{v}=-2$, $R=0,2,4$  as a function of the magnetic field $B$.
The legend indicates for each plot symbol which spin polarization the atomic cloud has and which $J'$-level is addressed in the REMPI measurement.
		The error bars correspond to the 95\% confidence intervals of the fits according to Eq.$\:$(\ref{eq:Lorentzian}). The dashed lines are calculations.
 The asterisk (*) next to a $m_F$ quantum number marks a two-atom scattering state.}
	\label{fig3}
\end{figure}
We now carry out a more quantitative analysis of the Zeeman energy level shifts. 
Since the excited $\text{v}'=66$, $A^1\Sigma_u^+$ level is essentially  insensitive to applied magnetic fields, the observed line shifts in Fig.\:\ref{fig2} correspond directly to the Zeeman shifts of the atomic states or the  molecular product states.
In order to determine the center positions $\nu_{i}$ of each observed resonance line $i$ we fit the function
\begin{equation}
\label{eq:Lorentzian}	
N(\nu)/N_0 = \exp \left( {\sum\limits_{i} \frac{-A_i}{(\nu-\nu_{i})^2+(\Gamma_i/2)^2}} \right)
\end{equation}
to the data of the remaining atom fraction, $N(\nu)/N_0$. Here,  $A_i$ is an amplitude, and $\Gamma_i$ is the linewidth.
 The results are presented in Fig.$\:$\ref{fig3}, together with
  additional REMPI data beyond the frequency range shown in
Fig.$\:$\ref{fig2}. The additional data correspond to  transitions from $\text{v}=-2,R=2$ and $R = 4$
 towards $J'=3$ of $\text{v}'=66$, $A^1\Sigma_u^+$ \cite{SelectionRule}. Together with these, our study covers all possible \cite{NotePossible} rotational product states $R=0,2,4$ of the $\text{v}=-2$ state.
 Figure\:\ref{fig3}  shows that the $J' = 1$ and $J' = 3$ data sets for the v = -2, $R = 2$ state lie precisely on top of each other, as required by consistency.

Next, we compare the measured energy positions of the product molecule levels to calculated Zeeman level positions (dashed lines).
Since the molecules are extremely weakly bound, their magnetic moment is essentially equal to the sum of the magnetic moments of the free atoms $a, b$.
In the limit of the linear Zeeman effect, the energy level positions are then given by
 $ E_b+g_f\cdot\mu_\text{B}\cdot(m_{f,a}+m_{f,b})\cdot B$, where $E_b$ is the binding energy of the respective state at zero field, $\mu_\text{B}$ is Bohr's magneton,  $g_f=-1/2$ is the $g$-factor, and $m_{f,a/b}$ is the magnetic quantum number of atom a/b, respectively.
Figure \ref{fig3}  clearly shows that three-body recombination produces $m_F = +2$ molecules in an ensemble of $m_f = +1$ atoms and it produces $m_F = -2$ molecules in an ensemble of $m_f = -1$ atoms.
Apparently, the total magnetic quantum number $m_F = m_{f,a} + m_{f,b}$  of the two atoms forming the molecule is conserved in three-body recombination.
This goes well along with a propensity rule for the conservation of the $F$ quantum number and the parity, observed in our previous work \cite{Wolf2017}.

\begin{figure}[b]
	\centering
	\includegraphics[width=\columnwidth]{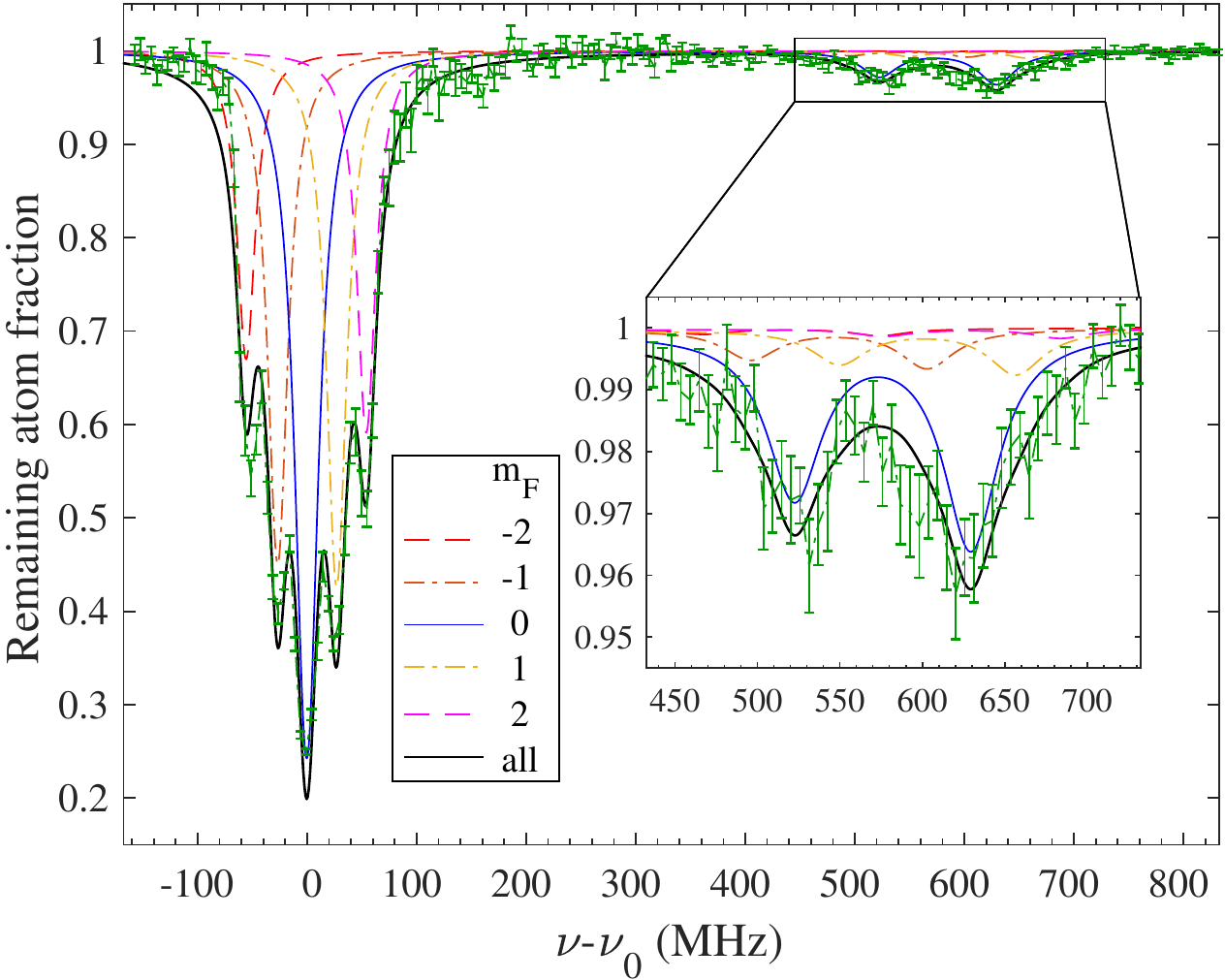}
	\caption{REMPI spectrum for an initial cloud of atoms prepared in a mix of $m_f$ quantum numbers at a magnetic field $B=37\:\text{G}$. The strong multi-line signal at $\nu-\nu_0 \approx 0$ corresponds to photoassociation. The signal is decomposed into the different  $m_F$ components for atom pairs
		using  Eq. (\ref{eq:Lorentzian}).
	  The inset is a zoom into the $\text{v}=-2,R=0$ and $\text{v}=-2,R=2$ molecular signals. Here, the dashed and solid colored lines correspond to calculations for the individual molecular $m_F$ populations which are formed in three-body recombination. The black solid line is the sum of these populations. }
	\label{fig4}
\end{figure}
Finally, we investigate molecular products in an atomic sample consisting of  a mix of $m_f$ states which has been produced by ramping the $B$-field through zero field. 
 Figure \ref{fig4} shows a REMPI spectrum for $B=37\:\text{G}$. As expected, the photoassociation signal around $\nu_0$ is now a multi-line structure involving all $m_F$ components $-2,-1,0,1,2$ resulting from all pair combinations of $m_f=-1,0,1$ atoms.
 The v$ = -2, R = 0, 2$ molecular signal between 450 MHz and 700 MHz (inset is a zoom) is now a broad double dip as a result of the ten overlapping molecular signals corresponding to the different combinations of $R$ and $m_F$ quantum numbers. Since no further substructure is observable a direct decomposition into molecular subcomponents is problematic. Therefore, we choose the following way for the analysis of the molecular signal. We first determine the three $m_f$-populations of the initial atom cloud and with these we predict the v$ = -2, R, m_F$  populations of the molecules and the corresponding spectrum which can be compared to the measurements. 
 From additional photoassociation loss measurements which are carried out without ion trap and a simple
 photoassociation model which assumes  the same photoassociation efficiency for all $m_F=m_{f,a}+m_{f,b}$  atom pairs, we extract the following $m_f$ distribution:   25\%, 45\%, 30\%  for $m_f=-1, 0, 1 $, respectively. 
Using this, we  predict the three-body recombination product molecule distribution to be 9\%, 17\%, 41\%, 20\%, 13\% for $m_F = -2, -1, 0, 1, 2$, respectively. Here, we use that the $m_F$ quantum number is conserved for three-body recombination and assume that the three-body recombination rate constant is identical for the different $m_F$ channels.
 The corresponding calculated $m_F$ signals are shown in the inset of Fig.~\ref{fig4}, using individual linewidths of $40\:$MHz. 
 The black line which is the sum over all signals agrees quite well with  experimentally observed double-dip structure.

In conclusion, we have carried out state-to-state measurements for three-body recombination with neutral $^{87}$Rb atoms in the regime of non-resonant two-particle interactions. We have demonstrated spectroscopic resolution of  the $m_F$ magnetic quantum states of the product molecules, in addition to the previously resolved vibrational, rotational, hyperfine degrees of freedom. Our studies suggest the propensity rule that the $m_F$ quantum number is  conserved when forming weakly-bound Rb$_2$ molecules via three-body recombination of cold $^{87}$Rb atoms.

For the future, we plan to probe more deeply bound molecular states  in order to investigate whether this and the previously found  propensity rules change, as indicated by our earlier work \cite{Haerter2013}.
From such experiments we can possibly gain information on how much the short range details of the interaction potentials influence the product distribution. Furthermore, it will be interesting to test whether the propensity rules found for $^{87}$Rb break down for other atomic species. $^{87}$Rb is special since its triplet and singlet scattering lengths are almost identical. 
Finally, we plan to further extend the detection method to also resolve the
magnetic quantum numbers $m_R$ of molecular mechanical rotation, as this would complete the experimental method for determining all internal quantum numbers of the molecule for the case that $\vec R$ and $\vec F$ are decoupled. 

We would like to thank Shinsuke Haze, Paul Julienne, Jos\'{e} D'Incao, and Eberhard Tiemann for fruitful discussions.
This work was supported by the German Research Foundation (DFG, Deutsche Forschungsgemeinschaft) within contract 399903135 and by the Baden-W\"{u}rttemberg Stiftung through the Internationale Spitzenforschung program (contract BWST\_ISF2017-061). M.D. acknowledges support from Universit\"{a}t Ulm and Ulmer Universit\"{a}tsgesellschaft (UUG) through a Forschungsbonus grant.

\end{document}